# Spectral Tuning of Polarization Selective Reflections Bands in GLAD deposited HfAlN chiral sculptured thin films

Samiran Bairagi, Marcus Lorentzon, Firat Angay, Roger Magnusson, Jens Birch, Ching-Lien Hsiao, Naureen Ghafoor*, Kenneth Järrendahl*

*Thin film physics division, Department of Physics, Chemistry & Biology, Linköping University, Linköping, Sweden SE-58334*

*Coressponding Author, Naureen Ghafoor: naureen.ghafoor@liu.se, Kenneth Järrendahl: kenneth.jarrendahl@liu.se

*Abstract*

We present the first report on fabrication of Hafnium aluminum nitride chiral sculptured thin films (CSTFs) using reactive magnetron sputtering in a glancing angle deposition configuration, and the analysis of its optical polarization properties. The resulting CSTFs were designed to give interference extrema or so-called circular Bragg (CB) resonances at desired wavelengths in the region from 370 to 690 nm. This was achieved by tailoring the growth of the chiral thin films to obtain a dielectric pitch between 87 and 260.9 nm. The spectral positions of the obtained CB resonances were compared to values from analytical expressions. Contrary to the common case where the dielectric pitch is half of the growth-related rotational pitch due to a 180° symmetry, this pitch was shown to be the same as the rotational pitch. It is concluded that this is due to the *c*-axis of the CSTF being tilted about 45° from the substrate normal. The morphology and crystallographic characterizations were done using scanning electron microscopy and X-ray diffraction, respectively, while the tilt of the crystal lattice was corroborated using X-ray diffraction pole figures. The optical response from the CSTFs was analyzed using Mueller matrix spectroscopic ellipsometry from which the degree of circular polarization at the CB resonances was obtained. In addition, a strong non-reciprocal reflection was observed which could be attributed to the helicoidal morphology and the intrinsic crystal tilt. An optical layered model of the chiral structure including azimuthal twist and using the Cauchy dispersion relations was used to simulate the Mueller matrix elements and compare with the ellipsometry measurements. The correlation between the simulated and experimental data gave information of the morphological parameters of the CSTF and its optical properties.

# 1. INTRODUCTION

Sculptured thin films (STFs) have garnered significant interest in the advancement of modern optics and nanophotonics, offering a distinctive approach to modulate the propagation and interaction of light with nanostructured surfaces. In this regard, Chiral sculptured thin films (CSTFs), which are a subset of STFs, stand out due to their intricate helical morphology. This allows them to intrinsically exhibit handedness-selective optical properties without requiring external surface alignment forces, thereby offering stability against environmental factors like pressure and temperature changes [1]. This handedness-selectivity is often termed as the circular Bragg (CB) phenomenon and facilitates reflection (or transmission) of circularly polarized light of the same (or opposite) handedness depending on handedness of the chiral medium, provided the medium is adequately thick with a sufficiently large number of periods[2]. This unique behavior allows for the development of novel applications in diverse areas, for example, advanced polarization filters, bioabsorption devices, optoelectronic devices, reflective displays, laser mirrors, photovoltaics, microcavities, and biological sensors[1], [3], [4].

Several approaches have previously been made to accomplish this selective reflection, such as quaterwave plates with linear polarizers, Fresnel rhombs, and cholesteric liquid crystals[5] but they suffer from structural limitations such as lateral broadening with growth[4], unstable structure with internal stresses[5], low deposition rates or sometimes due to changes in temperature and pressure[6]. In contrast, physical vapor deposition (PVD) grown CSTFs offer much more reliability thanks to the flexibility of the technique, enabling structural and compositional modifications of the chiral medium. This allows for strategies to optimize the functional characteristics of CSTFs such as tuning the optical response by modulating deposition parameters in magnetron sputtered glancing angle deposition (GLAD) or adjusting the underlying material properties, whether it be shifting the circularly polarized reflection bands or broadening their range. While sputter deposition techniques offer high precision, they are not without limitations. Long deposition periods can lead to column broadening, which in turn affects the optical properties of the resulting CSTF. This constraint poses a significant challenge for large-scale production and commercial applications.

In this research, we enhance the advantages of PVD CSTFs by introducing ternary nitride based materials, more specifically, hafnium aluminium nitride (HfAlN), which provide a combination of high refractive index, low optical absorption, chemical stability, and mechanical durability, making them highly suitable for the fabrication of robust and functional CSTFs. Furthermore, they are compatible with sputter deposition techniques, enabling precise control over film thickness and morphology. We will also address the limitations of sputter growth by presenting the first report to our knowledge for spectral tuning of polarization-selective reflection bands in HfAlN CSTFs fabricated using GLAD. Specifically, we show that by tilting the crystallographic *c*-axis of the sample, it is possible to obtain a longer dielectric pitch equal to the rotational pitch itself, thereby allowing shorter deposition times and improving the possibilities to fabricate samples which exhibit circular Bragg phenomena at longer wavelengths. This method provides an alternative approach for tuning the circular Bragg optical characteristics of CSTFs, thus extending their applicability and efficiency in various optical components. The interplay between the helicoidal morphology and the intrinsic tilted structure also results in exiting non-reciprocal reflection properties. By combining insights from the structure-property relationships, optical modelling and simulations, and the effect of different substrates and tilts of *c*-axis on the optical characteristics of HfAlN CSTFs, this work aims to advance the understanding of exhibition of circular Bragg phenomena. In a longer perspective it can lay the groundwork for their implementation in optical and optoelectronic devices.

## 2. EXPERIMENTAL

Chiral sculptured thin films (CSTFs) of HfAlN with a helical morphology were synthesized by direct current (DC) reactive magnetron sputter deposition in an ultra-high vacuum chamber, equipped with multi-movement manipulators on the sample stage and magnetrons (Mantis Deposition Ltd., Oxfordshire, UK). The system was evacuated below $4 \times 10^{-6}$ Pa before depositions. Hf and Al targets (99.999% purity, 7.62 cm diameter and 0.3 cm thick) were used as the sputtering source, and $2 \times 2$ cm$^2$ 111-oriented Si with a native oxide was used as the substrate. Some depositions were also made using Corning 7059 glass substrates. Before deposition, the silicon substrates were cleaned in an ultrasonic bath, sequentially employing isopropanol and acetone. After five minutes of cleaning, high purity $N_2$ gas was used to dry the substrates. Inside the chamber, the Hf and Al targets were sputter-cleaned at low powers for ten minutes in an Ar environment prior to the film deposition process. The main sputter deposition procedure was conducted in an ambient atmosphere consisting of a mixture of Ar and $N_2$ gases (99.99999% purity). The partial pressure ratio of this combination was maintained at 2:1, and the total working pressure was maintained at 0.4 Pa. The substrate was heated and maintained at 573 K during depositions and powers of 20 W and 120 W were applied to Hf and Al sputter sources, respectively. The distance between the substrate and the sources was set to 10 cm and the sample holder was aligned such that the incoming particle flux arrived at the substrate at an incident angle of $\alpha = 80°$ (the deposition angle) from the substrate normal, while the sources themselves were positioned opposite to each other, on the same side. A schematic view of the deposition setup and the resulting morphology is shown in Fig. 1.

The CSTFs used in this study have left-handed morphology, ensured by the anti-clockwise substrate rotation according to Fig. 1a. The rotational pitch $d_p$, i.e., the distance related to one complete (360°) rotation, was chosen to obtain main circular Bragg (CB) resonance within the near-ultraviolet (NUV) to near-infrared (NIR) wavelength region. Stepwise rotation was employed during the film deposition, where the substrate was positioned at a glancing deposition angle α, and the growth of a segment (a layer) occurred over a pre-set time, $t_s$. After each $t_s$, the substrate was rotated by an angle of $\xi = 60°$, completing a full rotation after six segments and the time $6\ t_s$. In this way the ratio of rotation speed to deposition rate, dictated $d_\mathrm{p}$. The depositions were conducted for approximately 6 hours to achieve a total film thickness between 1.7 to 2.8 μm. Some 100 nm thick HfAlN thin films were also grown on Si with similar conditions but without the GLAD configuration. These isotropic films were used for initial determination of the average refractive index and absorption onset for HfAlN.

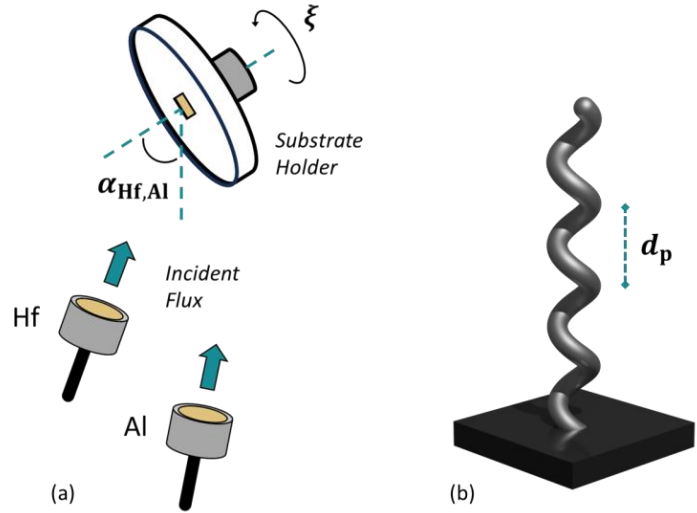

Fig. 1: Schematics of (a) the deposition setup defining the positive twist-angle $\xi$ giving (b) a left-handed helicoidal structure with rotational pitch $d_p$.

The morphological study of the helical thin films was performed with a Zeiss LEO 1550 field-emission scanning electron microscope (FE-SEM, Zeiss, Oberkochen, Germany), set at a 3 kV accelerating voltage and using an InLens detector. Side view examinations were facilitated through cleaved samples, providing insights into the number of periods, pitch, and tilt of each segment making up the CSTF, as well as total thickness of the film. The top view was used to assess the film surface. Crystal structure and pole figures were obtained using Philips X'Pert-MRD operated with Cu Ka radiation. The primary optics were crossed slits ($2 \times 2$ mm$^2$) and the secondary optics consisted of a parallel plate collimator (0.27°) with a flat graphite crystal monochromator. Pole figure measurements were performed along $\psi$ (tilt) and $\phi$ (rotation) directions in the range 0 to 80° and 0 to 360°, respectively, with 5° step sizes at a collection rate of 0.88 s/step. The microstructure of the samples was analyzed using transmission electron microscopy (TEM). Cross sectional specimens were prepared by conventional mechanical polishing followed by Ar-ion etching. The imaging was performed using an FEI Tecnai G2 TF 20 UT field-emission instrument operated at 200 keV for a point resolution of 0.19 nm. To elucidate the optical polarization properties, a comprehensive Mueller matrix analysis of the samples was performed. Samples were measured using a dual rotating compensator ellipsometer (RC2) and analyzed using the CompleteEase software package (both from J. A. Woollam Co. Inc., Lincoln, NE, USA). Incidence angles for all measurements ranged from 20° to 80° in steps of 10°, over a spectral range of 210 - 1690 nm. Each sample underwent a full 360° rotation during measurement, enabling the collection of 72 Mueller matrices in steps of 5° azimuthal angle, yielding all 16 elements $m_{ij}$ (i, j = 1, 2, 3, 4) of the normalized Mueller matrix. In addition, other important parameters such as degree of polarization and reflected intensities of left- and right-handed circularly polarized light from the sample were obtained from the measurements.

## 3. THEORY

### 3.1. The Stokes-Mueller formalism

To understand the optical properties of the chiral sculptured thin film (CSTF) samples used in this study, we utilize the Mueller matrix formalism thanks to its capability to effectively describe anisotropic properties and also address partially polarizing samples. This 4x4 matrix can be used to describe any optical element and how it affects light interacting with it according to $\mathbf{S}_o = \mathbf{M}\mathbf{S}_i$, or explicitly:

$$\begin{bmatrix} I \\ Q \\ U \\ V \end{bmatrix}_o = \begin{bmatrix} 1 & m_{12} & m_{13} & m_{14} \\ m_{21} & m_{22} & m_{23} & m_{24} \\ m_{31} & m_{32} & m_{33} & m_{34} \\ m_{41} & m_{42} & m_{43} & m_{44} \end{bmatrix} \begin{bmatrix} I \\ Q \\ U \\ V \end{bmatrix}_i \qquad (1)$$

Here the description is restricted to normalized Mueller matrices ($m_{11}$ = 1). The reflected ($\mathbf{S}_o$) and incident ($\mathbf{S}_i$) light are described by Stokes vectors according to:

$$\mathbf{S} = \begin{bmatrix} I \\ Q \\ U \\ V \end{bmatrix} = \begin{bmatrix} I_x + I_y \\ I_x - I_y \\ I_{+45} + I_{-45} \\ I_R - I_L \end{bmatrix} \qquad (2)$$

Where, $I_x, I_y, I_{+45}$ and $I_{-45}$ are irradiances of linearly polarized light along $x$, $y$, $+45°$ and $-45°$ directions and $I_R$ and $I_L$ are irradiances of right- and left-handed circular polarizations, respectively. The elements $m_{41}$ and $m_{14}$ are of special interest for this study as a positive / negative $m_{41}$ (or $m_{14}$) value relates to the degree of right/left-handed circular polarization, respectively[7]. For incident unpolarized light, the degree of polarization $P$, and the degree of circular polarization $P_c$ of the reflected light can be related to the Mueller matrix elements according to:

$$P = \sqrt{m_{21} + m_{31} + m_{41}} \qquad (3)$$

and

$$P_c = m_{41} \qquad (4)$$

### 3.2. The circular Bragg phenomena

The periodically rotated structures studied in this work demonstrate reflection of elliptically polarized light with one handedness and simultanous transmission of the other handedness. As an analogue to reflection from other periodic media, especially of x-rays from crystalline materials, this is referred to as a Bragg phenomenon, and the range in which it takes place the Bragg regime. When the propagation vector of light with a high degree of circular polarization, $P_c$, matches the structure of the helicoidal structure, it fulfills a Bragg resonance condition. The circular Bragg (CB) resonances will occur as extrema in optical data from the helicoidal media, and are studied in spectral plots of reflectance, transmittance, or Mueller matrix elements. Due to the use of mainly absorbing substrates in this study, we will focus on reflected light.

Although wave interaction with helicoidal media can be calculated giving, for example, reflectance and transmittance for full wavlength ranges[8] we can predict positions of the CB resonaces from basic interference considerations[9]. We first assume an ideal case with normally incident light on a layered structure where each layer (analogous with the segment introduced above) is an anisotropic but non-dispersive medium with in-plane refractive indices $n_a$ and $n_b$. Subsequent layers are twisted $\xi$ degrees from each other and after $N_\ell = \frac{360°}{\xi}$ number of twists, a full turn, corresponding to the rotational pitch $d_\mathrm{p}$, is achieved. This pitch is directly connected to the actual rotation of the sample stage during growth. The thickness of a layer will be $d_\ell = \frac{d_\mathrm{p}}{N_\ell}$. If we consider small enough $\xi$, making the chiral structure approximately continous, the CB resonances would be positioned at the wavelength:

$$\lambda_0 = 2\Omega_0 \cdot n_\mathrm{av} \tag{5}$$

where $\Omega_0$ is the dielectric pitch and $n_\mathrm{av} = \frac{n_a + n_b}{2}$ is the average refractive index of the layers. It is by far most common that the helicoidal media has a half-turn (180°) symmetry resulting in a dielectric pitch which is half of the rotational pitch, $\Omega_0 = \frac{d_\mathrm{p}}{2}$. However, as will be demonstrated in this study it is possible to obtain media with a dielectric pitch equal to the rotational pitch, $\Omega_0 = d_\mathrm{p}$. The handedness of the resulting reflected light depends upon the handedness of the chiral sample. We define that a sample with left-handed chiral morphology according to Fig. 1 will reflect left-handed circularly polarized light, and transmit right-handed circularly polarized light. To account for different orders of CB resonances that occur when shorter wavelengths fulfill the interference conditions in periodic media Eq. 5 is modified according to:

$$\lambda_{0m} = \frac{2\Omega_0 \cdot n_\mathrm{av}}{m} \cos\theta'; m = (1,2,3,\ldots) \tag{6}$$

where, $m$ denotes the order of the resonance. Angular dependence is also taken into account in Eq. 6 where $\theta'$ is the refraction angle related to the angle of incidence $\theta$ through the law of refraction. Higher order resonances ($m = 2,3,\ldots$) do not occour for ideal helicoidal media in the case of normal incidence ($\theta = \theta' = 0°$), only for oblique angles[2]. For oblique incidence the refractive index $n_\mathrm{c}$ of the third axis must be considered when calculating $n_\mathrm{av}$. For $m = 1$ and normal incidence the expression Eq. 5 will be obtained ($\lambda_{01} \equiv \lambda_0$). The corresponding full width at half maxima (spectral width) of the main CB resonance can be obtained from:

$$\Delta\lambda_0 = 2\Omega_0 \cdot \Delta n \cos\theta' \tag{7}$$

where $\Delta n = |n_a - n_b|$ is the linear birefringence exhibited by the sample.

Helicoidal morphology can give rise to additional resonances. If we increase the twist angle $\xi$ the structure can eventually not be treated as continous and due to the discrete steps shorter dielectric pitches $\Omega_p$ in the helicoidal arrangement also cause CB resonances This is referred to as ambichirality[9] and the expected resonances can be described by altering Eq. 6 according to:

$$\lambda_{qm} = \frac{2\Omega_q \cdot n_{av}}{m} \cos\theta'; \; q = (0, \pm 1, \pm 2, \ldots); \; m = (1,2,3,\ldots) \tag{8}$$

with the are ambichiral dielectric pitches described by:

$$\Omega_q = \frac{\Omega_0}{1+q\eta}; \; q = (0, \pm 1, \pm 2, \ldots) \tag{9}$$

where $\eta = \frac{\Omega_0}{d_\ell} = \Omega_0 \frac{N_\ell}{d_p}$, i.e. the number of layers in the main dielectric pitch. Setting, $q = 0$ we get the expression corresponding to the resonances at $\lambda_{0m}$ defined in Eq. 6. For positive values of the integer ($q = 1,2,3,\ldots$) the ambichiral resonances will be associated with reflections of the same handedness as the main resonance, and for negative values ($q = -1, -2, -3$) the corresponding handedness will be the opposite. For the typical pitches and refractive indeces related to this work the ambichiral resonances were usually appearing outside the measured wavelength range below 200 nm.

## 4. RESULTS AND DISCUSSION

### 4.1. Structure and morphology

Scanning electron microscopy images of the chiral sculptured thin film (CSTF) samples used in this study are shown in Fig. 2. They were grown with the objective to obtain the main circular Bragg (CB) resonances $\lambda_0$ at different wavelengths in the NUV-NIR region. The chosen segment deposition times $t_s$ from 131 to 420 seconds and the resulting film thicknesses are presented in Table 1 together with other relevant data estimated from the SEM micrographs. Considering a limited total deposition time shorter $t_s$ allowed for a larger number of rotational pitches. Typically, the density or porosity in GLAD deposited thin films is controlled by varying the angle of incoming flux ($\alpha$) reaching the substrate[10]. However, it is observed that even when $\alpha$ (and all the other parameters) are kept the same for all samples, the CSTFs grew denser with a more coalesced morphology, as $t_s$ was reduced. This is because shorter times resulted in growth of shorter segments at each subsequent step angle making up the CSTF. This diminished the ballistic shadowing effect at each step and hence the overall density of the samples.

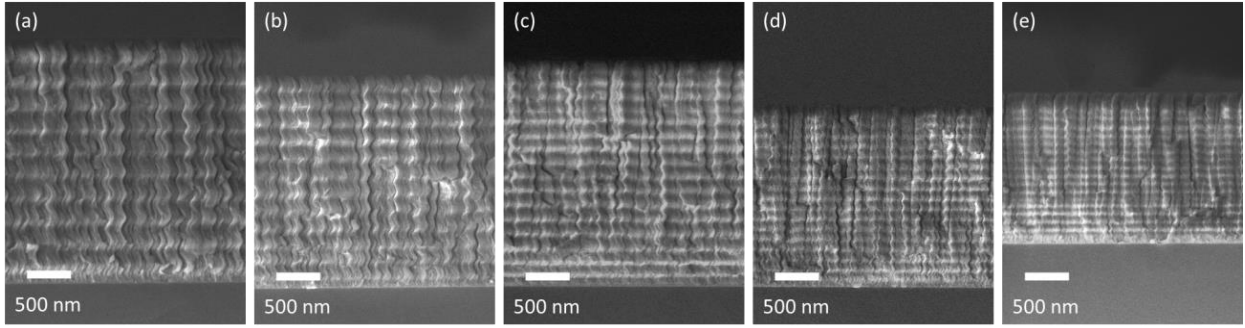

*Fig. 2: Cross sectional SEM images of samples (a) A, (b) B, (c) C, (d) D, and (e).*

In Table 1 the measured tilts of the individual segments in the CSTFs are presented and determined to have an average value of ≈ 61° from the base of the substrate (that is, ≈ 29° from the substrate normal). Note that because of a short period of residue of adatoms after terminating the deposition continued growth can occur with the result that $N_P$ is not always a whole number. Top-view SEM images for all samples reveal a similar coalesced, layered morphology as observed in our previous work for samples grown at 0.4 Pa working pressure and hence is not discussed further[11]. With the uniform helical morphology observed in all samples, we showcase the possibility of fabricating ternary nitride CSTFs with several micrometer thicknesses using reactive magnetron sputtering.

*Table 1: Summary of measured morphological sample parameters. The number of pitches were calculated from $d_f/d_p$.*

| Sample | A | B | C | D | E |
|---|---|---|---|---|---|
| Segment deposition time, $t_s$ (s) | 420 | 300 | 192 | 175 | 131 |
| Film thickness, $d_f$ (nm) | 2807 | 2431 | 2583 | 2057.5 | 1698 |
| Rotational pitch, $d_p$ (nm) | 260.9 | 202.6 | 147.6 | 111.2 | 87.1 |
| Segment tilt (°) | 59.4 | 61.1 | 63.1 | 63.4 | 59.5 |
| Number of rotational pitches, $N_p$ | 10.8 | 12 | 17.5 | 18.5 | 19.5 |

## 4.2. Crystallographic orientation

We established in our previous study[11] that at lower working pressures, GLAD deposited thin films resulted in a biaxial crystallographic texture, with preferential inclination of the *c*-axis towards the direction of incident flux from the sputter source. In particular, it was observed that the *c*-axis of the sample grown at 3 mTorr working pressure was tilted ≈ 50° with respect to the substrate normal, regardless of the varying morphological tilts of the samples (resulting from growth at higher working pressure). A similar investigation was carried out in this study to determine if the HfAlN chiral sculptured thin films (CSTFs) exhibited a lattice tilt. This was performed using X-ray diffraction (XRD) pole figure measurements by mapping the tilt (from 0° to 85°) and orientation (from 0° to 360°) of HfAlN {0002} and HfAlN {10$\bar{1}$0} at fixed 2Θ angles of 36.041° and 33.216°, respectively. The results are plotted in Fig. 3.

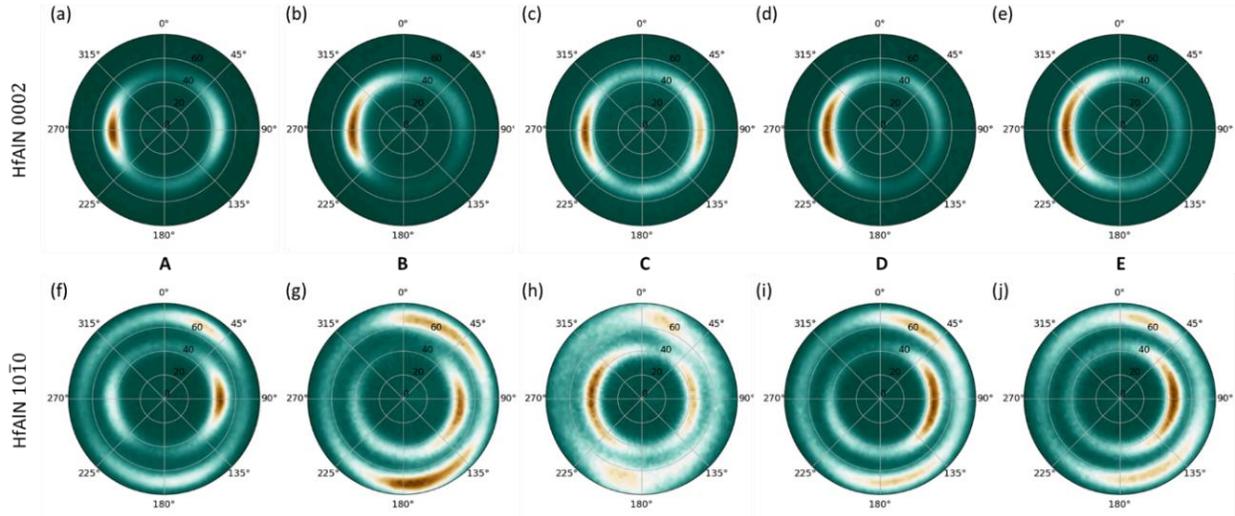

Fig. 3: XRD pole figures of all samples for (a-e) HfAlN 0002 and (f-j) HfAlN 10$\bar{1}$0.

In Fig. 3(a-e) it can be observed that all samples exhibit the presence of a strong pole located at $\psi$ ≈ 42°, 43°, 46°, 45° and 45° for samples A, B, C, D and E, respectively. This confirms the c-axis HfAlN {0002} in all samples is tilted with respect to the substrate normal. Fig. 3(f-g) also shows presence of dominant poles located at $\phi$ ≈ 47°, 48°, 43°, 45° and 45° for samples A, B, C, D and E, respectively, indicating preferential orientation of the corresponding basal planes, HfAlN {10$\bar{1}$0}, with some degree of disorder. Looking at the angular spread of the poles in all samples it can be observed that while both HfAlN {0002} and {10$\bar{1}$0} exhibit an azimuthal variation of Δ$\psi$ ≈ 10°, their mosaic variation is quite large and ranges for {0002} from Δ$\phi$ ≈ 45° in samples A and B to Δ$\phi$ ≈ 90° in samples D and E for {0002}. For {10$\bar{1}$0} the range is from Δ$\phi$ ≈ 120° in samples A and B to Δ$\phi$ ≈ 220° in samples D and E. This means that regardless of the helical morphology, the CSTF samples in this study exhibit a biaxial texture (with some degree of random orientation in HfAlN {10$\bar{1}$0}) where the c-planes effectively remain inclined by 45° with respect to the substrate normal, but with a 27% variation in their azimuthal orientation around the c-axis. This information is illustrated in Fig. 4, where Fig. 4a and b show a generalized representation of HfAlN {0002} and {10$\bar{1}$0} pole figure data. We hypothesize that during the early stages of deposition process, the directionality in incoming flux creates a template for the preferential orientation of HfAlN lattice planes, as early as the growth of the first step of the CSTF. But due to the short deposition time at each step and rotation of the substrate thereafter, this preferential orientation does not develop until significant growth of the film has taken place. This is why we see a ring-like feature of weaker intensity in the poles in Fig. 3, showing preferential orientation with some degree of randomness. Discussions related to the presence of weaker intensity poles (i) on the opposite side of the dominant poles in Fig. 3(a-e), and (ii) due to the other prismatic planes $\{1\bar{1}00\}$ and $\{01\bar{1}0\}$ of the hexagonal HfAlN crystal in Fig. 3(f-j) can be found elsewhere (Section 4.2 in [11]).

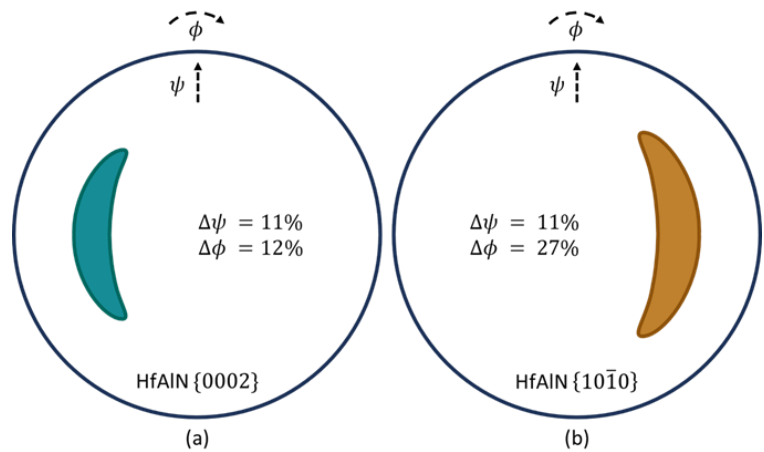

Fig. 4: Schematic illustration of HfAlN pole figures showing variation and spatial orientation of (a) HfAlN $\{0002\}$ and (b) HfAlN $\{10\bar{1}0\}$ poles.

### 4.3. Microstructural analysis

To assess the tilt of the crystal lattice and understand its local orientation, sample B was further characterized using TEM. An overview of the cross-sectional STEM-HAADF image in Fig. 5a confirmed a dense uniform helical morphology of CSTFs with 12 periods and structural pitch ≈ 200 nm, in agreement with the SEM investigations. Since CSTFs exhibited some in-plane randomness, tracing the individual chiral morphology in cross-sectional view was limited in the sample prepared by conventional polishing. However, the initial growth, diameter and crystal orientation in the pitches can be observed from the regions of bright contrast of an overlaid dark field image in Fig 5a. It showed that the helical nanostructure (spiral) attained a diameter of ≈ 60 nm after growth of the first pitch, and its subsequent growth occurred uniformly with same crystal orientation. This crystal orientation was further investigated for 180° rotated regions (marked 1 and 2) from the same pitch of the spiral shown in bright field TEM image in Fig. 5b. High resolution TEM images in Fig. 5c, 5d, and 5e (lattice resolved image with FFT as inset ) revealed that growth of the spiral occurred uniformly with 0002 lattice planes tilted by ≈ 43° with respect to the substrate normal, independent of substrate rotation angle. These results were in agreement with XRD pole figure analysis, and electron diffraction measurements (not shown here), which confirmed the in-plane fiber texture of the films.

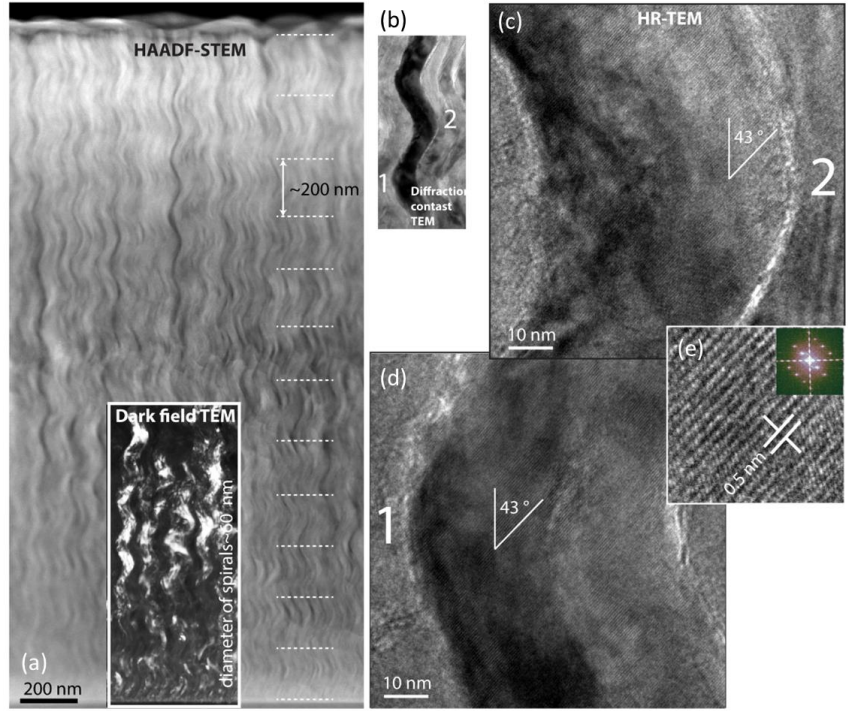

*Fig. 5: TEM analysis of sample B with 12 periods. (a) Overview STEM-HAADF image showing 12 structural pitches of ~200 nm thickness. Inset in the image shows overlayed dark-field image, (b) diffraction contrast bright field TEM of a segment of a chiral structure, (c) and (d) are magnified 1 and 2 regions of segments shown in (b). (e) is the lattice resolved image and corresponding FFT showing lattice tilt and spacing in regions 1 and 2.*

### 4.4. Circular Bragg resonances

Initially an optical characterization of the isotropic HfAlN thin films grown without the GLAD condition ($\alpha < 20°$) was made. This gave an average index of $n_{av} \approx 1.95$ and the onset of absorption below 420 nm, which were used as initial values for the optical analysis presented in this section.

Mueller matrix spectroscopic ellipsometry (MMSE) data for the different fabricated CSTFs are presented in Fig. 6. The normalized m41 element was primarily chosen to elucidate the associated degree of circular polarization (Eq. 4) of the reflected light when the incoming light is unpolarized. In Fig. 6 the $m_{41}$ data measured at a fixed azimuthal angle, and a 20° angle of incidence is plotted as a function of wavelength.

When examining MMSE data, it was noted that the CB resonances appear less pronounced than in ideal cases. Additionally, the measured MMSE data was found to vary with the azimuthal angle. This helped in identifying the CB resonances, as analyzing the data across all 72 rotational directions provided valuable insight to distinguish the extrema arising due to the helicoidal films from the thickness fringes. For the five samples A, B, C, D, and E, the main CB resonances (marked in Fig. 6) are found at approximately $\lambda_0 = 950$, 770, 625, 450 and 370 nm, respectively. Even though the extrema are suppressed, it is clear that the main CB resonance is positioned at different wavelengths in line with the different $t_s$ used for the growth. A simulation of the CB resonances (based on the optical model described in section 4.6) for the samples with varying pitches is shown in supplementary Fig.S. 1.

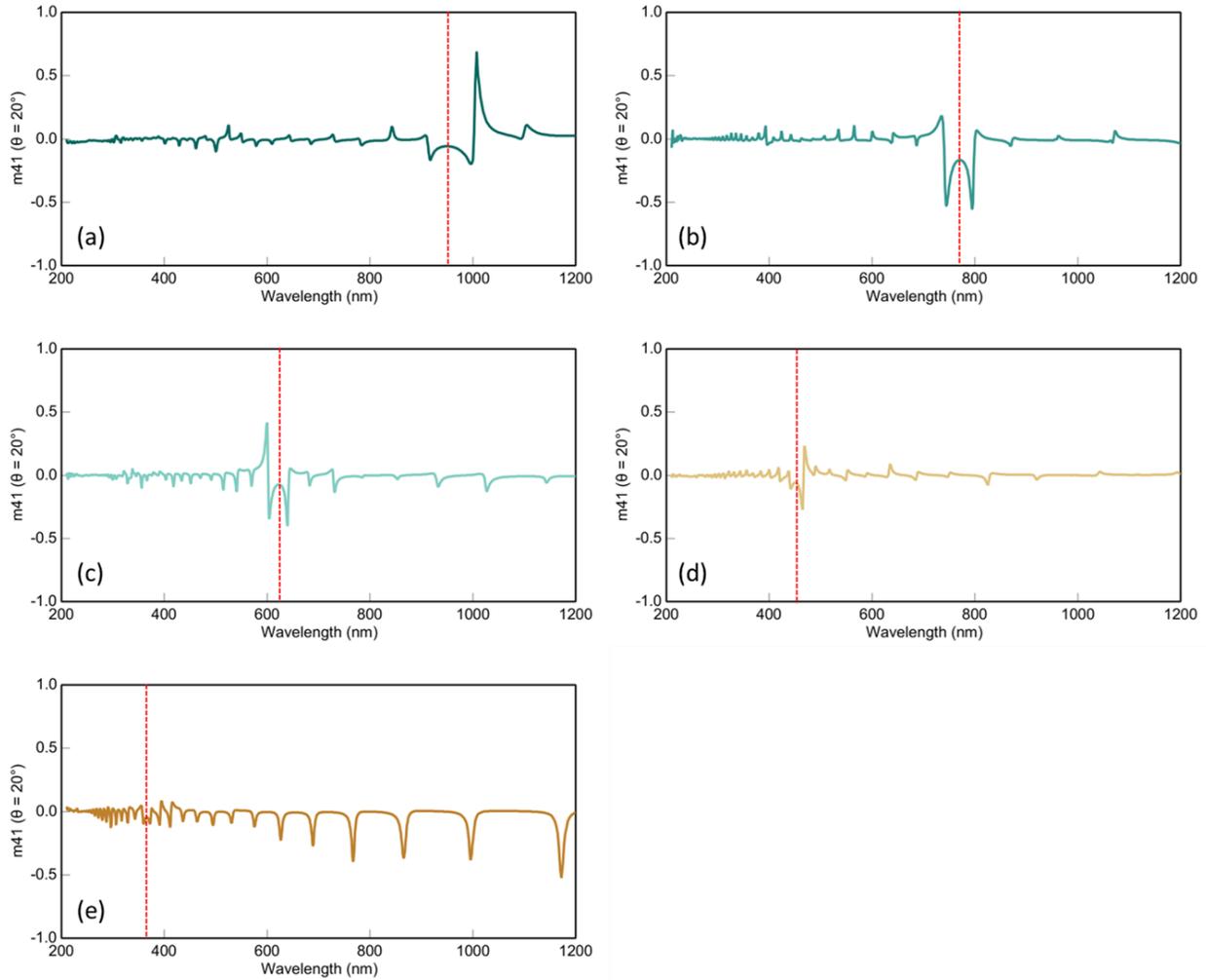

Fig. 6: The Mueller matrix element $m_{41}$ versus $\lambda$ for samples (a) A, (b) B, (c) C, (d) D, and (e) E.

There are several combined causes for the suppression of the CB extrema. First, spectral features from the helicoidal films may be affected by interference due to the total film thickness which can disturb the CB resonances depending on the optical properties of the surrounding media. More explicitly, the relation between the refractive indices of the ambient, film, and substrate may cause a phase shift making thickness fringe minima overlap with CB maxima, and vice versa. From simulations (supplementary Fig.S. 2) we observe that in such cases we get the characteristic "folded" resonance features seen in Fig. 6. Moreover, it is understood from simulations (supplementary Fig.S. 3) and literature[5], [12] that the number of rotational pitches is important and that $N_\mathrm{p} \gtrsim 40$ is required to exhibit strong circular Bragg phenomena in CSTFs. In our case we have $N_\mathrm{p} < 20$, due to time restrictions for the sputter growth. Another related fact influencing the spectra is the limited growth precision mentioned above resulting in non-integer number of rotations. We can also note that cases where the CB extrema are affected by thickness fringes will be especially affected by low and non-integer $N_\mathrm{p}$.

To address the influence of the refractive index of the surrounding media and see if the "folded" features seen in Fig. 6 and Fig.S. 2 could be avoided, we made additional film deposition on glass instead of Si. That is, changing the substrate refractive index from above to below the average refractive index of the film. Data for sample B2, made with the same growth parameters as sample B, is presented in Fig. 7. The resulting CB resonances are still rather weak but are more Gaussian-shaped without the folded appearance seen in the films grown on Si. The main CB resonance is positioned at about 860 nm, i.e., at a higher value compared to sample B. This difference can be attributed to the difference in growth mode of the CSTF due to the choice of a different substrate.

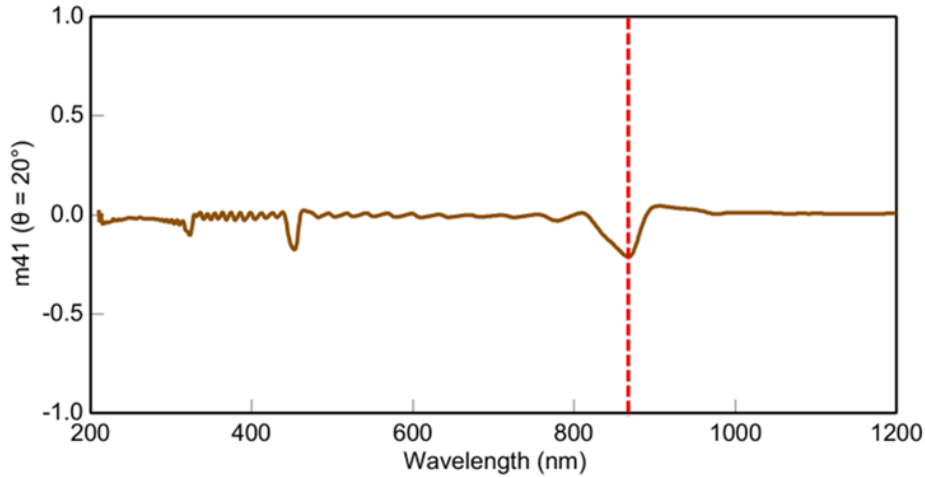

Fig. 7: The Mueller matrix element $m_{41}$ versus $\lambda$ for sample B2.

Several other imperfections can be associated with CSTFs taking them further from the ideal case with perfect homogenous spirals, which also results in less prominent extrema. For instance, in the present growth scheme, the moderately low growth temperature and low working pressure resulted in CSTFs with non-uniform size and shape. When comparing the resonances for the five samples, the features for sample D and especially sample E have the lowest magnitudes even though they have the highest number of rotational pitches. Since these samples show main resonances at lower wavelengths (450 and 370 nm, respectively) the absorption edge near 420 nm can be a probable cause to the weak resonances.

Along with the values for the main CB resonances, some related data are presented in Table 2. The initial value of the refractive index of $n \approx 1.95$ combined with the observed main resonance values impose that the dielectric pitch $\Omega_0$ could not be equal to half of the rotational pitch $d_p$ which is the common case, but is *equal to* the rotational pitch. That is, the grown helicoidal films have a 360°, not 180° rotational symmetry. By assuming that $\Omega_0 = d_p$ the average refractive index $n_{av}$, and linear birefringence $\Delta n$ of the samples can be determined from the observed spectral positions and the broadening of the main CB extrema using Eq. 6 and Eq. 7. In the approximative calculations $m = 0$, and $\theta' = \frac{\theta}{2} = 10°$ was used. The resulting $n_{av}$, and $\Delta n$ is presented in Table 2. As $t_s$, and thus $d_p$ is reduced, the samples from A to E exhibit increasing $n_{av}$, while $\Delta n$ decreases. The first is in line with typical normal dispersion for transparent materials but the different density and absorption in the samples, as discussed above, may influence the values.

Table 2: Summary of the main circular Bragg resonances and related optical data.

| Sample | A | B | C | D | E |
|---|---|---|---|---|---|
| $\lambda_0$ (nm) | 950 | 770 | 625 | 450 | 370 |
| $\lambda_0$ (eV) | 1.31 | 1.61 | 1.99 | 2.76 | 3.35 |
| $\Delta\lambda_0$ (nm) | 80 | 50 | 35 | 21 | 12 |
| $n_{av}$ | 1.85 | 1.93 | 2.15 | 2.05 | 2.16 |
| $\Delta n$ | 0.16 | 0.13 | 0.12 | 0.10 | 0.07 |

It has been shown that optical rotation is approximately proportional to the square of the linear birefringence[9]. Similar influence is expected for the present CB resonances where the different $t_s$ results in a reduced (local) birefringence. This is in line with the decreasing values for samples A to E in Table 2.

According to the expressions in section 3.2, other resonances would also occur in the spectra. From the obtained optical data and main dielectric pitches, ambichiral resonaces ($q = 1,2,...$) should occur at about 190 nm and below, outside the meaured wavelength region. That is, any additional spectral features should be ascribed to higher orders of the main resonance ($q = 0, m = 1,2,..$). Although even weaker than the main resonances they can be observed for the samples having the largest dielectric pitches and thus have their higher order resonances within the measured wavelength region. Higher order resonances are observed in Fig. 5 at approximative wavelengths according to: 488 nm for sample A, 400 nm for sample B, 450 nm and 320 for sample B2. From Eq. 6 (using $q = 0$) we can attribute these resonances to being of higher order with $m = 2$, and also $m = 3$ for sample B2. By observing the MMSE-data for varying azimuthal angles it is possible to observe additional resonances according to. $m = 3$ and 4 for sample A, $m = 3$ for sample B, $m = 4$ for sample B, and $m = 2$ for sample C.

### 4.5. Non-reciprocal reflection

An obvious application of CSTFs are their ability to selectively reflect (or transmit) left-handed circularly polarized light (LCP) or right-handed circularly polarized light (RCP), depending upon the handedness of the CSTF[2], [13]. This can be well elucidated by observing individual detected intensities (irradiances) of the left- and right-handed components of the circularly polarized reflected light, for unpolarized incident light. Fig. 8 shows the normalized m41 element for sample B. It is the same data as in Fig. 6b with the main CB resonance at 770 nm but here presented as a function of wavenumber $(1/\lambda)$ making the thickness interference fringes evenly spaced (in this and the next section all data is plotted versus wavenumber). Fig. 8 also shows the detected intensity of LCP and RCP, and it can be observed that the sample reflects a higher degree of LCP light compared to RCP showcasing the handedness selective behavior of the sample. This is also evident from the $m_{41}$ data, which is the difference between the LCP and RCP intensities, and results in an overall negative (or left-handed) degree of circular polarization for the reflected light.

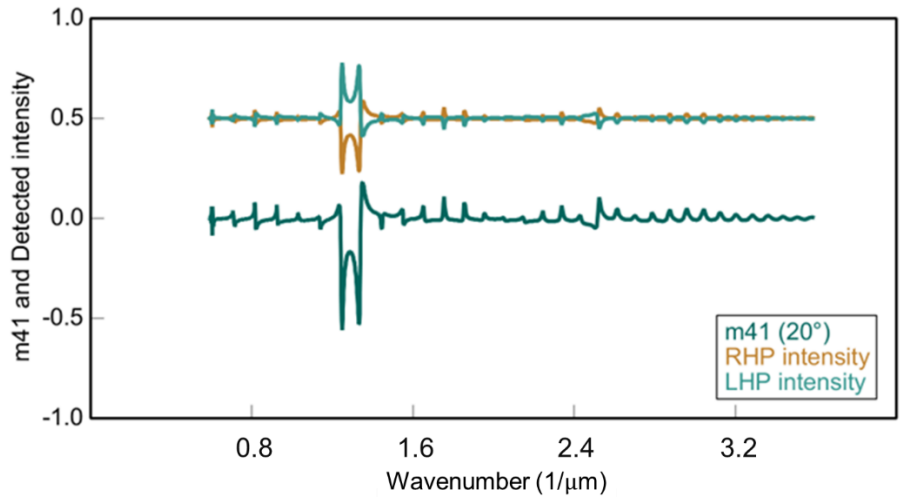

Fig. 8: Normalized m41 for sample B. Left- and right- handed detected intensities for unpolarized incident light.

In section 4.4 we concluded that our MMSE data is dependent on the azimuthal measurement angle. A more detailed study of the variation shows that a 180° azimuthal rotation results in flipped spectra, showing the opposite handedness. That is, a sample being left-handed in one direction becomes right-handed in the opposite direction in the region of the CB resonance. This is demonstrated in Fig. 9 for sample C, where the azimuthal direction used in Fig. 6c shows a left-handed resonance, whereas a rotation of 180° show the opposite. In Fig. 6 and 7 the direction showing most clear left-handedness was presented, but in all cases the opposite direction (rotated 180°) could have been chosen.

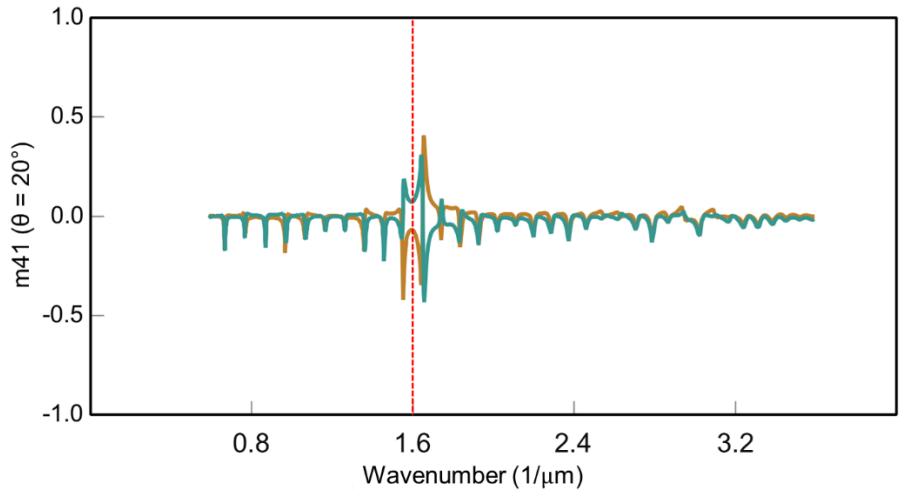

Fig. 9: Normalized $m_{41}$ for sample C, at two azimuthal angles differing 180°.

The azimuthal dependence implies that the fabricated helicoidal samples have non-reciprocal reflection properties. This can be studied by observing relations between the Mueller matrix elements, in particular $m_{14}$ and $m_{41}$, where $m_{14} \neq m_{41}$ in a non-reciprocal case. The relation to the 180° azimuthal difference showcased in Fig. 9 becomes evident in Fig. 10 where the two elements for the same sample (C) are plotted.

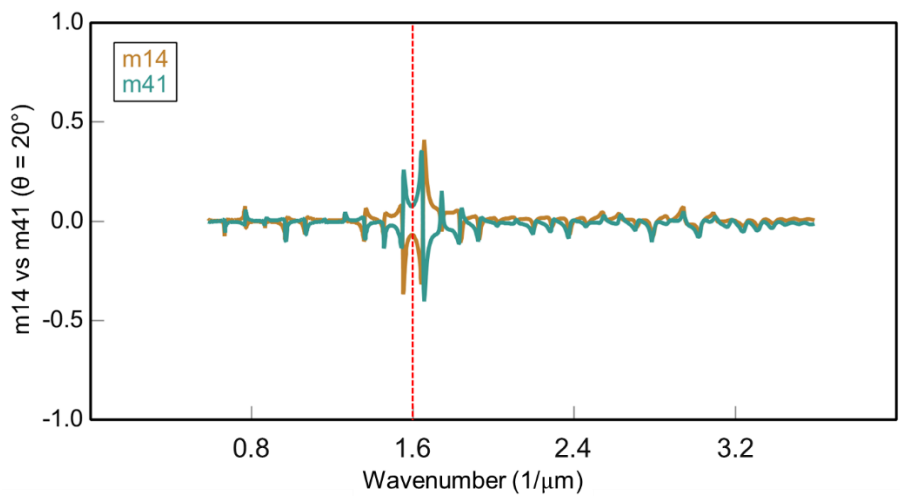

Fig. 10: Normalized $m_{14}$ and $m_{41}$ for sample C.

So far, we have observed that our CSTFs have 360° dielectric symmetry ($\Omega_0 = d_\text{p}$), azimuthal reflection dependence, and non-reciprocal reflection properties. These can be explained from the structure and morphology results discussed in section 4.1.

A key distinction between traditional CSTFs and the samples in this study is the tilted c-axis with respect to the substrate normal. Even when the chiral nanostructure grows almost parallel to the substrate normal, the HfAlN (0002) remains inclined towards the source by about 45° – due to this, the sample exhibits the interesting non-reciprocal optical behavior. Further insights into the non-reciprocity can be gained from Fig. 11a-c showing polar contour plots of the $m_{41}$ element where for three angles of incidence (20°, 40°, and 60°), $1/\lambda$ is plotted along the radius, and the azimuthal rotation is plotted along the circumference. The horizontal black dotted line indicates the position of the main CB resonance. It can be observed by the arcs in the plot, that around the CB resonance the azimuthal symmetry is two-fold (180°) as mentioned above, but in other regions the symmetry is four-fold (90°). This symmetry becomes more clear for higher angles of incidence. We corroborated the symmetry conditions by investigating the detected intensities of LCP and RCP at different azimuthal rotations for incoming unpolarized light. In Fig. 11d-f plots of $m_{14}$ vs $m_{41}$ are presented to show the non-reciprocity related to the polar plots.

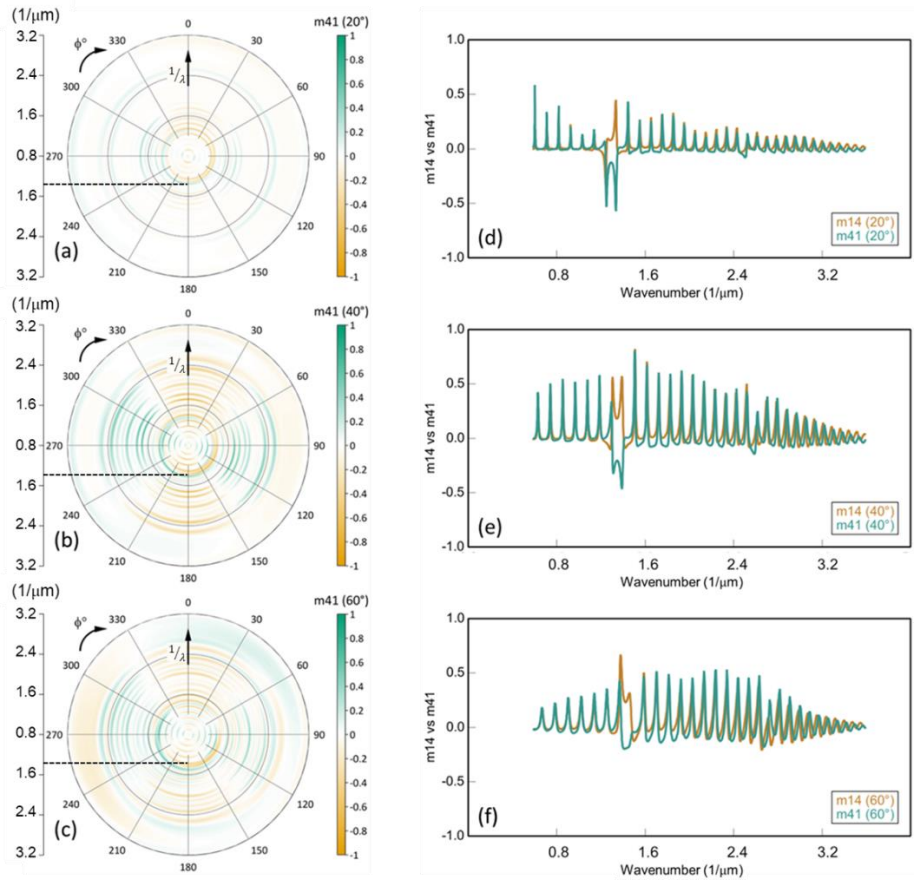

Fig. 11: (a-c) Contour plots and (d-f) $m_{14}$ and $m_{41}$ data showing non-reciprocal behaviour in sample B.

Based on this discussion, it can be assumed that the optical axis is dependent on the intrinsic orientation of the c-plane of the material (i.e., HfAlN [0002]). Typically, CSTFs are amorphous and do not have a c-axis or preferred growth orientation and in cases when CSTFs are crystalline, the c-axis is parallel to the substrate normal. Consequently, in these cases the optical properties of the sample remain identical regardless of azimuthal orientation. As a result, the dielectric pitch is typically half the rotational pitch. In our samples however, due to the tilted c-axis, one complete rotation of the structure is the same as one complete rotation of the optical axis. As a result, the dielectric pitch in our samples will be identical with the rotational pitch resulting in main CB resonances at twice as high wavelengths compared to the typical case. This can be seen as an advantage since a high degree of circular polarization can be achieved in a specific wavelength region with shorter pitches. That is, for the same $d_p$, the main CB regime can be obtained at twice the spectral positions compared to traditional CSTFs. The effect of the tilted crystal lattice on the spectral position of the CB resonances is shown in supplementary Fig.S. 4, where the Euler tilt angle θ is varied in a series of simulations (based on the optical model in section 4.6).

### 4.6. Optical modelling

The structural and morphological information of the CSTF samples in this study were extracted using SEM and XRD in section 4.1 and 4.2. Some of the obtained parameters are used here as input in a multi-layered optical model, which is employed to grasp the validity and dependence of the physical parameters on the optical properties. Given the complex nature of the samples and limitations of a one-dimensional model, only sample B was used for the optical modelling procedure.

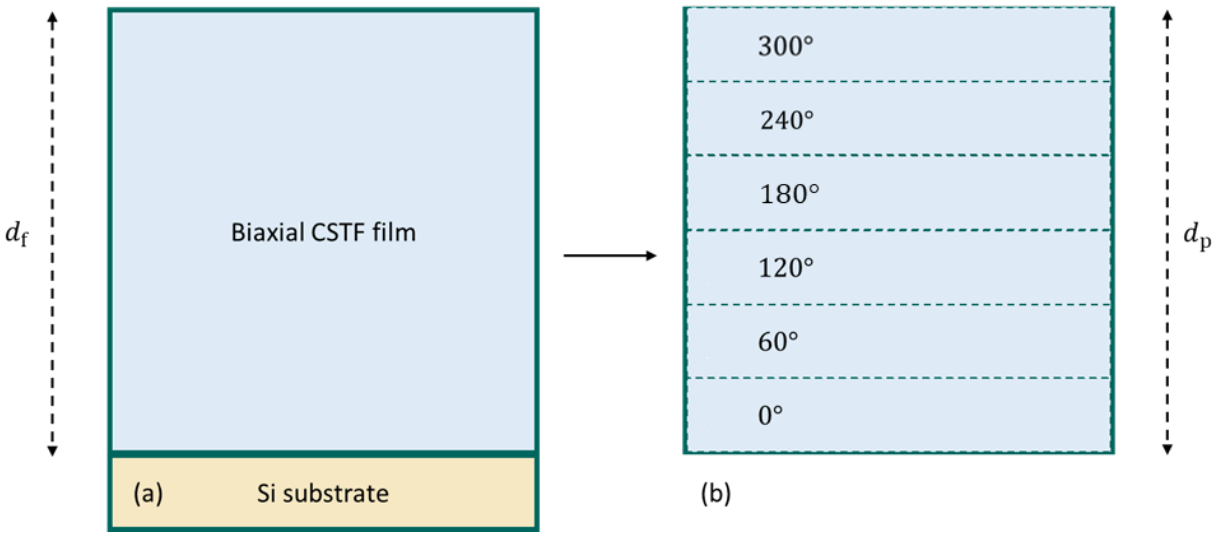

Fig. 12: Schematic of the optical model for (a) the entire CSTF sample consisting of multiple number of rotational pitches $N_p$ and (b) the sublayers making up one rotational pitch.

A schematic view of the optical model is shown in Fig. 12 and the simulations were made with the following considerations.

- Due to the helical and columnar structure, a real sample would exhibit local biaxial symmetry making the overall anisotropy rather complicated. Therefore, for each sublayer in the model we apply in-plane ($n_x$, $n_y$) and out-of-plane ($n_z$), refractive indices, resulting in the $z$-axis parallel to the substrate normal. The anisotropy is, however, further affected by the Euler angle tilt described below.

- The periodic structure (with 360° turns) of the CSTF, can be effectively built up by defining one rotational pitch and stacking $N_p$ copies of this. The total film thickness will then be $d_f = N_p \cdot d_p$, with values taken from Table 2.

- The *c*-axis of the wurtzite HfAlN crystal is inclined by ≈ 45° from the substrate normal. This tilt is included in the model by tilting the optical axis in each sublayer by setting the Euler angle tilt = 45°.

- The lateral composition gradient producing the intrinsic anisotropic properties (due to the complementary placement of the Hf and Al sputter sources), will rotate around the substrate normal during growth. This is included in the model by the azimuthal Euler angle. The 60° twist of substrate rotation is modelled by setting the azimuthal angle from 0° to 300° in steps of 60° in the six sublayers.

- Non - idealities such as surface roughness, void density, and variations in helicoidal attributes are neglected and the CSTFs are represented as a uniform and homogeneous chiral film, without relying on effective medium theories.

- For the fitting procedure the wavelength spectrum was restricted in the range 496 – 1690 nm (0.73 – 2.5 eV) at a 20° angle of incidence.

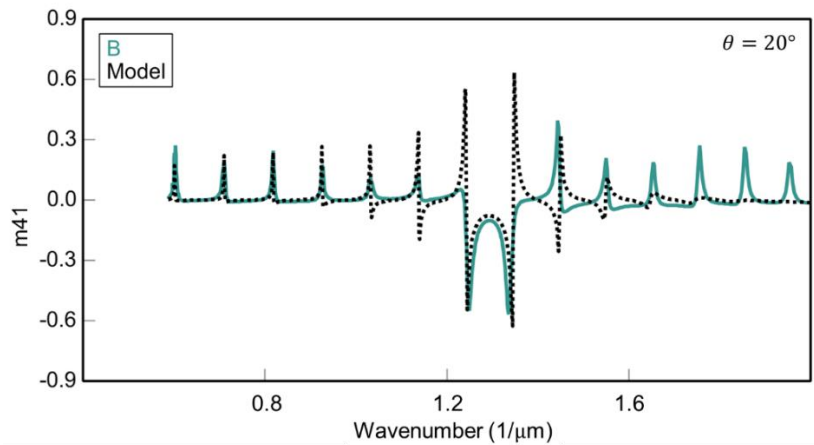

Fig. 13: Experimental and modelled fit of $m_{41}$ data for sample B, with experimental data in solid green and generated model data in dashed black.

The biaxial layer optical properties of the CSTF were modelled by parameterizing them as a function of wavelength using a Cauchy dispersion relation as $n_m = A_m + B_m/\lambda^2 + C_m/\lambda^4$ for each of the three directions $m \in \{x, y, z\}$ [14], [15]. This optical model was used to fit all elements of the Mueller matrix spectra simultaneously using regression analysis with $d_p, A_m, B_m, C_m$, and the two Euler angles as fit parameters.

Fig. 13 shows an example of the best fit of the optical model to the experimental $m_{41}$ data for sample B (other matrix elements in Supplementary information). The values obtained from the fitting procedure agree closely with those determined in Section 4.1 and 4.3. The rotational pitch became $d_p$ = 205.43 nm whereas the measured value from SEM was 202.6 nm. The Euler tilt angle became 55.3° compared to the XRD pole value for the c-axis tilt was 44.6°. The resulting average index is $n_{av}$ = 1.88 (at 600 nm). This value was found close to the isotropic $n$ for a 58 nm $Hf_{0.16}A_{0.54}N_{0.30}$ thin film by Selvakumar et al. (in Fig. 4d of [16]), also grown using magnetron sputtering but with a different material composition compared to the films in this work.

Fig. 14 shows the complex refractive index in the three directions extracted from the optical model. While this study is the first to report optical properties of biaxially textured HfAlN CSTFs, the results are comparable to crystalline InAlN CSTFs grown using reactive magnetron sputtering [14]. Fig. 14 also includes optical data of the homogenous isotropic HfAlN thin film for comparison. This film was also modelled with a Cauchy dispersion resulting in a refractive index ($n_{iso}$) according to the dashed line and this average index $n_{av} \approx 1.95$ was used as an initial value for the analysis in section 4.4. The overall lower $n_{av}$ of the CSTF sample in comparison to $n_{iso}$ for the homogeneous thin film is plausible considering the porosity, which is a characteristic of GLAD deposited thin films.

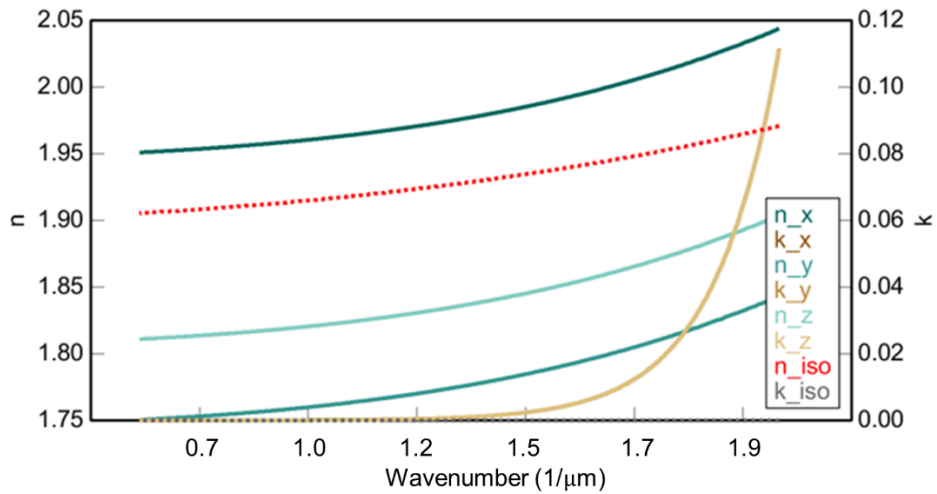

Fig. 14: Optical constants of HfAlN (Sample B) sample and an isotropic thin film

## 5. CONCLUSION

We present the first report on fabrication of Hafnium aluminum nitride chiral sculptured thin films (HfAlN CSTFs) and the analysis of its optical properties using Mueller matrix spectroscopic ellipsometry. The morphology and crystallographic characterizations were obtained from scanning electron microscopy and x-ray diffraction, respectively, while a 45° tilt of the crystal lattice was corroborated using x-ray diffraction pole figures. The resulting CSTFs were designed to exhibit circular Bragg (CB) phenomena at specific wavelengths 370 to 690 nm. This was done by tailoring the reactive magnetron sputtering GLAD depositions to get dielectric pitches between 87 and 260.9 nm. Although several imperfections causing suppression of the CB-resonances we could conclude that the spectral positions of the obtained main and higher order CB extrema were in agreement with analytical expressions. Interestingly, the resulting resonances gave that the rotational and dielectric pitch were the same, implying a 360° symmetry, not the typical 180° symmetry seen in helicoidal media. This could be explained by the *c*-axis tilt. An azimuthal dependence of the Mueller matrix spectra and a clear difference between the $m_{14}$ and $m_{41}$ revealed that the samples exhibited a strong non-reciprocal reflectance. A conceptual optical model using the Cauchy dispersion formalism was used to elucidate and simulate the correlation between morphological parameters of the CSTF and its resulting optical properties.

## 6. CREDIT STATEMENT

Conceptualization: S.B., R.M., C.-L.H. and K.J.; methodology: S.B., C.-L.H. and K.J.; resources: J.B., N.G., C.-L.H. and K.J.; data curation: S.B., F.A., M.L. and N.G.; writing—original draft preparation: S.B., R.M., and K.J; writing—review and editing: S.B., R.M., N.G., J.B. and K.J.; project administration: N.G. and K.J.; funding acquisition: J.B., N.G. and K.J. All authors have contributed to the writing and agreed to the published version of the manuscript.

## 8. SUPPLEMENTARY FIGURES

The following supplementary images were generated using data from sample B and the optical model discussed in section 4.6. The position of the main CB resonance in each case is marked using a dashed red line.

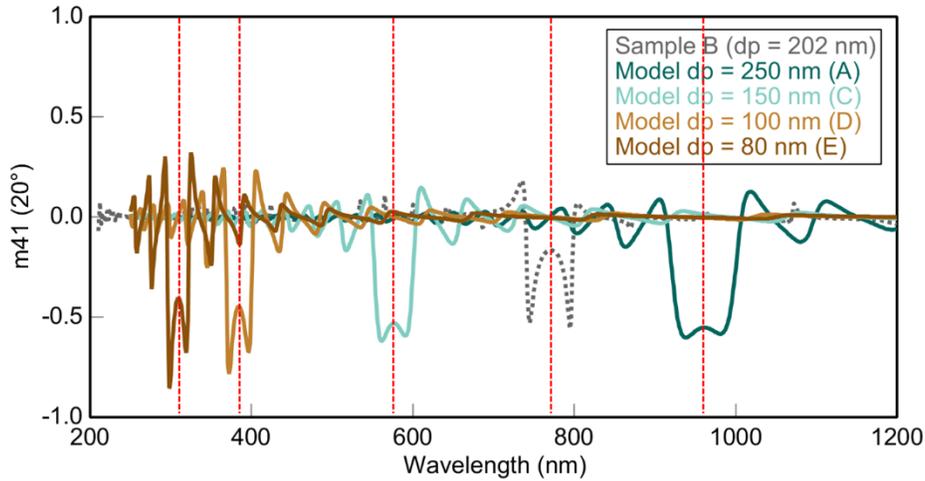

*Fig.S. 1: Effect of $d_p$ on the spectral position of CB resonance. This was used to estimate the $d_p$ for the deposition of samples A,B,C,D,E in section 4.1.*

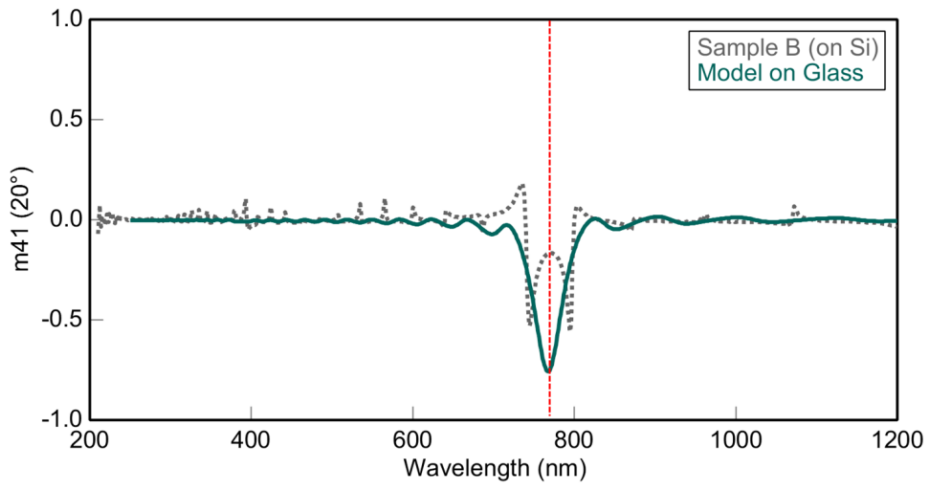

*Fig.S. 2: Effect of choice of substrate on the shape and intensity of the CB resonance. Notice the "folded" resonance when using a Si substrate ($n_{Si} > n_{av}$) compared to using a glass substrate ($n_{glass} \lesssim n_{av}$).*

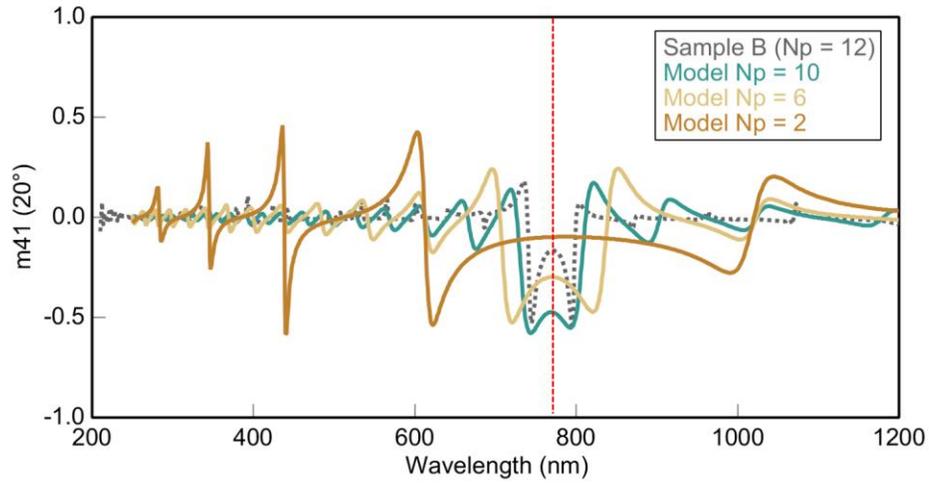

*Fig.S. 3: Effect of $N_p$ on the shape of CB resonance.*

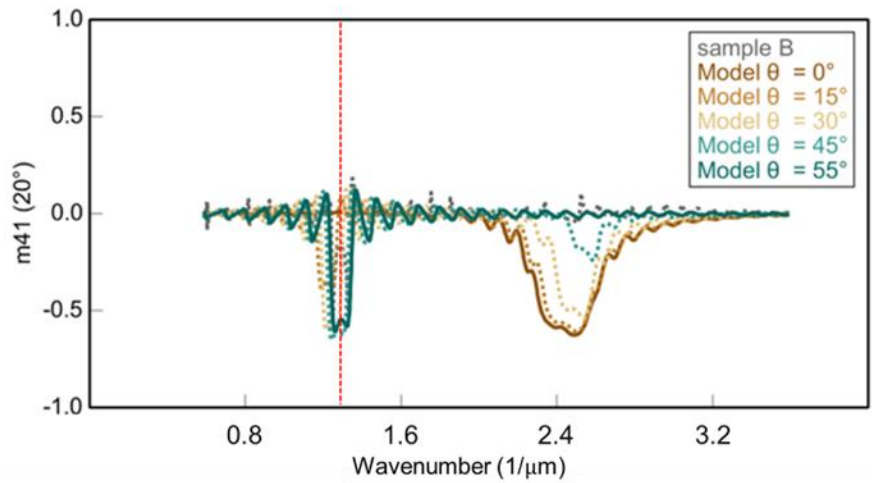

*Fig.S. 4: Effect of c-axis tilt on CB resonance by varying Euler tilt angle θ from 0° and 55°.*